\documentclass{article}
\usepackage{amssymb}
\usepackage{tikz}

\begin{document}

\title{Stability of stationary solutions in models of the Calvin cycle}

\author{Stefan Disselnk\"otter and Alan D. Rendall\\
Institut f\"ur Mathematik\\
Johannes Gutenberg-Universit\"at\\
Staudingerweg 9\\
D-55099 Mainz\\
Germany}

\date{}

\maketitle

\begin{abstract}
In this paper results are obtained concerning the number of positive 
stationary solutions in simple models of the Calvin cycle of photosynthesis
and the stability of these solutions. It is proved that there are open sets
of parameters in the model of Zhu et. al. \cite{zhu09} for which there exist 
two positive stationary solutions. There are never more than two isolated 
positive stationary solutions but under certain explicit special conditions 
on the parameters there is a whole continuum of positive stationary solutions. 
It is also shown that in the set of parameter values for which two isolated 
positive stationary solutions exist there is an open subset where one of the 
solutions is asymptotically stable and the other is unstable. In related 
models derived from the work of \cite{grimbs11}, for which it was known that 
more that one positive stationary solution exists, it is proved that there are 
parameter values for which one of these solutions is asymptotically stable and 
the other unstable. A key technical aspect of the proofs is to exploit the fact 
that there is a bifurcation where the centre manifold is one-dimensional.  

\end{abstract}

\section{Introduction}

The Calvin cycle is an important part of photosynthesis and many mathematical 
models have been proposed to describe it \cite{arnold11}, \cite{jablonsky11}.
These vary widely in the number of chemical species included and the 
kinetics chosen for the individual reactions. In what follows we concentrate
on some of the simplest models with the aim of obtaining rigorous results on
the number of stationary solutions of the models and their stability. It may be 
hoped that a deeper understanding of the simpler models will lead to new 
approaches to analysing the more comprehensive ones. 

In \cite{zhu09} a model of the Calvin cycle was introduced which is 
a system of ordinary differential equations describing the concentrations of
five substances. This level of biological detail is similar to that found in
the standard textbook \cite{alberts02} on cell biology. Based on computer
modelling the authors of \cite{zhu09} conclude that their system has only one
steady state for fixed values of the parameters under certain biological 
restrictions. (A stationary solution of 
a system of ODE is one which is independent of time and a steady state is
an alternative name for a stationary solution.) They explictly exclude the
case of non-isolated steady states from consideration. They do not make general
statements about the stability of the steady states although some results of
simulations included in the paper indicate the stability of the steady state 
considered. In what follows we prove that under certain explicit restrictions 
on the parameters of the system a continuum of positive steady states occurs. 
We also give a proof that when these restrictions are not satisfied there 
exist at most two positive steady states for each choice of the parameters.
It is shown that there do exist open sets of parameters for which there are
two positive steady states. This does not contradict the results of 
\cite{zhu09} since the biologically motivated fixed choice of Michaelis
constants made there excludes the cases where more than one steady state is
present in the model.

It is proved that there are parameters for which an asymptotically stable 
positive steady state exists. Thus for these parameters any solution 
which has the property that at some time the concentrations are sufficiently 
close to those in the steady state converges to the steady state at late times.
There are also open regions in parameter space 
for which no positive steady state exists and regions for which the only 
positive steady state is unstable. The stability of the state where all 
concentrations are zero is also dependent on the parameters. Thus for some 
choices of parameters there are solutions for which all concentrations tend to 
zero at late times and for other choices of parameters no solutions of this 
kind exist. More detailed versions of these statements can be found in Theorem 1
and Theorem 2 in Sect. 2. It follows from Theorem 4 of \cite{rendall14} that 
there are runaway solutions of this model where all concentrations tend to 
infinity at late times. 

In \cite{grimbs11} the authors introduced what looks at first sight like a 
small modification of the model of \cite{zhu09} by rescaling two of the 
coefficients. It turns out, however, that this modifies the dynamics 
significantly. These authors considered different possibilities for the 
kinetics. The model of \cite{grimbs11} with Michaelis-Menten kinetics is the 
main subject of Sect. 3. It was already known from \cite{grimbs11} and 
\cite{rendall14} that this model has two positive steady states for certain 
choices of the parameters. However nothing had been proved about their 
stability. Here we show that there are
parameters for which one of these steady states is stable and one unstable.
It is also shown that this implies analogous statements for a more complicated
model, also introduced in \cite{grimbs11}, where each basic reaction is 
described using a Michaelis-Menten scheme with a substrate, an enzyme and a 
substrate-enzyme complex. Details are given in Theorem 3. 

Sect. 4 is concerned with equations derived from a model introduced in
\cite{grimbs11} where the concentration of ATP is included as an additional
variable and the diffusion of ATP is taken into account. This leads to a 
system of reaction-diffusion equations. Setting the diffusion coefficient to 
zero in this model or restricting consideration to spatially homogeneous 
solutions gives rise to a system of ODE which was called the MAdh system in 
\cite{rendall14}. It was proved there that there are parameters for which this 
model has two positive steady states. However once again nothing was proved 
about the stability of these solutions. Here we show that for certain values 
of the parameters one of the steady states is stable and the other unstable.
Details are in Theorem 4. The last section contains a summary of the results
of the paper and an outlook on possible future developments. 

\section{The model of Zhu et. al.}

This section is concerned with a model of the Calvin cycle introduced by
Zhu et. al. \cite{zhu09}. The basic system of equations is
\begin{eqnarray}
&&\frac{dx_{\rm RuBP}}{dt}=v_5-v_1,\label{zhu1}\\
&&\frac{dx_{\rm PGA}}{dt}=2v_1-v_2-v_6,\label{zhu2}\\
&&\frac{dx_{\rm DPGA}}{dt}=v_2-v_3,\label{zhu3}\\
&&\frac{dx_{\rm GAP}}{dt}=v_3-v_4-v_7,\label{zhu4}\\
&&\frac{dx_{\rm Ru5P}}{dt}=\frac35 v_4-v_5\label{zhu5}.
\end{eqnarray}
Here $x_X$ denotes the concentration of the substance X and the substances 
involved are ribulose bisphosphate (RuBP), phosphoglycerate (PGA),
diphosphoglycerate (DPGA), glyceraldehyde phosphate (GAP) and ribulose
5-phosphate (Ru5P). The $v_i$ are reaction rates and are given by the 
following expressions of Michaelis-Menten type.
\begin{eqnarray}
&&v_1=\frac{k_1x_{\rm RuBP}}{x_{\rm RuBP}+K_{m1}},\label{rr1}\\
&&v_2=\frac{k_2x_{\rm ATP}x_{\rm PGA}}{(x_{\rm PGA}+K_{m21})(x_{\rm ATP}+K_{m22})}
,\label{rr2}\\
&&v_3=\frac{k_3x_{\rm DPGA}}{x_{\rm DPGA}+K_{m3}},\label{rr3}\\
&&v_4=\frac{k_4x_{\rm GAP}}{x_{\rm GAP}+K_{m4}},\label{rr4}\\
&&v_5=\frac{k_5x_{\rm ATP}x_{\rm Ru5P}}{(x_{\rm Ru5P}+K_{m51})(x_{\rm ATP}+K_{m52})},
\label{rr5}\\
&&v_6=\frac{k_6x_{\rm PGA}}{x_{\rm PGA}+K_{m6}},\label{rr6}\\
&&v_7=\frac{k_7x_{\rm GAP}}{x_{\rm GAP}+K_{m7}}.\label{rr7}
\end{eqnarray}
Here the $k_i$ are the maximal reaction rates and the $K_{mi}$ Michaelis 
constants. The concentration of adenosine triphosphate (ATP) is not modelled 
dynamically but taken to be maintained at a constant value. Notice that, 
as remarked in \cite{arnold11}, the expression for the sixth of these 
equations given in \cite{zhu09} is not correct and it has been modified 
accordingly here. While this change affects the biological interpretation of 
some of the parameters in the equations it does not change the mathematical 
properties of the model. In \cite{zhu09} the authors looked for positive 
steady states of this system using a computer program and they 
concluded that there exists at most one solution of this type for fixed 
values of the parameters if the concentrations and parameters are in
biologically relevant ranges. In what follows we investigate to what extent 
statements of this type can be proved analytically and what can be said about 
the stability of the steady states.

Steady states are characterized by the equations 
\begin{equation}\label{linearmmz}
v_1=v_5, 2v_1=v_2+v_6, v_2=v_3, v_3=v_4+v_7\ \ {\rm and}\ \ v_5=\frac35 v_4 
\end{equation}
for the reaction rates. Combining these gives 
\begin{equation}\label{combmmz}
\frac15 v_4-v_6-v_7=0.
\end{equation} 
It follows from (\ref{linearmmz}) and (\ref{combmmz}) that if 
$\beta=\frac{v_7}{v_4}$ then 
\begin{equation}\label{rational}
\frac{v_6}{v_2}=\frac{1-5\beta}{5(1+\beta)}.
\end{equation}
For a positive steady state we must have $0<\beta<\frac15$.  
It can be checked that the equations (\ref{linearmmz}) for steady 
states are equivalent to (\ref{combmmz}), (\ref{rational}) and 
the equations
\begin{equation}\label{remaining}
v_5=\frac35 v_4, v_3=v_4+v_7\ \ {\rm and}\ \ v_1=\frac35 v_4.
\end{equation}
Thus any set $(v_2,v_4,v_6,v_7)$ which solves (\ref{combmmz}) and 
(\ref{rational}) can be completed by means of (\ref{remaining}) to a solution 
of (\ref{linearmmz}). For any solution of (\ref{linearmmz}) there is at
most one set of concentrations which give rise to these $v_i$ and when
such concentrations exist they define a steady state of the 
system (\ref{zhu1})-(\ref{rr7}). For instance 
\begin{equation}
x_{\rm Ru5P}=\frac{v_5(x_{\rm ATP}+K_{m52})K_{m51}}{k_5x_{\rm ATP}-(x_{\rm ATP}+K_{m52})v_5}
\end{equation}
provided the denominator is positive and otherwise there is no positive 
steady state. For convenience we define a paramater 
$\kappa=(K_{m7}-K_{m4})(K_{m6}-K_{m21})$.

\noindent
{\bf Lemma 1} The system (\ref{zhu1})-(\ref{rr7}) with given positive 
parameters satisfying $\kappa\ne 0$ has at most two positive steady states. If 
$\kappa>0$ or precisely one of the factors in the product defining $\kappa$ is 
non-zero it has at most one positive steady state. 

\noindent
{\bf Proof} Suppose first that $\kappa\ne 0$. Then both
factors are non-zero. Now
\begin{eqnarray}
&&\frac{v_7}{v_4}=\frac{k_7}{k_4}\frac{x_{\rm GAP}+K_{m4}}{x_{\rm GAP}+K_{m7}}
\label{q47},\\
&&\frac{v_6}{v_2}=\frac{k_6(x_{\rm ATP}+K_{m22})}{k_2x_{\rm ATP}}
\frac{x_{\rm PGA}+K_{m21}}{x_{\rm PGA}+K_{m6}}\label{q26}.
\end{eqnarray}
It follows from (\ref{q47}) and $K_{m7}\ne K_{m4}$ that $k_7\ne k_4\beta$ and
from (\ref{q26}) and $K_{m6}\ne K_{m21}$
that $5k_6(1+\beta)(x_{\rm ATP}+K_{m22})\ne k_2(1-5\beta)x_{\rm ATP}$. Thus
\begin{eqnarray}
&&x_{\rm PGA}=\frac{k_2(1-5\beta)K_{m6}x_{\rm ATP}
-5k_6(1+\beta)K_{m21}(x_{\rm ATP}+K_{m22})}
{5k_6(1+\beta)(x_{\rm ATP}+K_{m22})-k_2(1-5\beta)x_{\rm ATP}},\label{pgaq}\\
&&x_{\rm GAP}=\frac{k_4\beta K_{m7}-k_7K_{m4}}{k_7-k_4\beta}.\label{gapq}
\end{eqnarray}
The expressions for $x_{\rm PGA}$ and 
$x_{\rm GAP}$ can be substituted back into the expressions for the reaction rates 
$v_4$ and $v_6$ to get
\begin{eqnarray}
&&v_4=\frac{k_4\beta K_{m7}-k_7K_{m4}}
{\beta(K_{m7}-K_{m4})},\\
&&v_6=\frac{k_2(1-5\beta)K_{m6}x_{\rm ATP}-5k_6(1+\beta)K_{m21}(x_{\rm ATP}+K_{m22})}
{5(1+\beta)(x_{\rm ATP}+K_{m22})(K_{m6}-K_{m21})}.
\end{eqnarray}
As a consequence of (\ref{combmmz}) we need that 
$v_6=\left(\frac15-\beta\right)v_4$ in order to get a steady state. This 
implies the vanishing of the cubic polynomial $p(\beta)$ given by
\begin{eqnarray}\label{cubic}
&(1-5\beta)(k_4\beta K_{m7}-k_7K_{m4})[(1+\beta)(x_{\rm ATP}+K_{m22})(K_{m6}-K_{m21})]
&\\
&-[k_2(1-5\beta)K_{m6}x_{\rm ATP}-5k_6(1+\beta)K_{m21}(x_{\rm ATP}+K_{m22})]
\beta(K_{m7}-K_{m4}).&\nonumber
\end{eqnarray}
The sign of $K_{m6}-K_{m21}$ is the same as that of 
$k_2(1-5\beta)K_{m6}x_{\rm ATP}-5k_6(1+\beta)K_{m21}(x_{\rm ATP}+K_{m22})$ 
since $v_6>0$. This is in turn the same as the sign of 
$5k_6(1+\beta)(x_{\rm ATP}+K_{m22})-k_2(1-5\beta)x_{\rm ATP}$ since 
$x_{\rm PGA}>0$. The sign of $K_{m7}-K_{m4}$ is the same as that of 
$k_4\beta K_{m7}-k_7K_{m4}$ since $v_4>0$. This is in turn the same as the sign 
of $k_7-k_4\beta$ since $x_{\rm GAP}>0$. 

The sign of the coefficient of the 
leading term in the polynomial $p$ is opposite to that of $K_{m6}-K_{m21}$.
The sign of $p(0)$ is also opposite to that of $K_{m6}-K_{m21}$. The sign of 
$p(1/5)$ is that of $K_{m7}-K_{m4}$. Under the assumptions of Lemma 1 the sign 
of $p(0)$ is opposite to that of $p$ for large negative values of its 
argument. Hence $p$ has at least one negative root and at most two positive 
ones. This gives the first conclusion of the lemma. If the signs of 
$K_{m7}-K_{m4}$ and $K_{m6}-K_{m21}$ are the same then $p$ changes sign 
in the interval $(0,1/5)$ and again for $\beta>1/5$. Thus it has precisely one
root in the interval $(0,1/5)$. Now consider the case that $K_{m7}-K_{m4}=0$.
Then $k_7-k_4\beta=0$ and $\beta$ is determined by the reaction constants.
Under the assumptions of Lemma 1 the inequality $K_{m6}-K_{m21}\ne 0$ holds in 
this case and so (\ref{pgaq}) holds. Thus $x_{\rm PGA}$, $v_2$ and $v_6$ are 
determined. The equation $v_6=\left(\frac15-\beta\right)v_4$ then determines 
$v_4$. This means that all reaction rates and all concentrations are 
determined. The proof in the case where $K_{m7}-K_{m4}\ne 0$ and 
$K_{m6}-K_{m21}=0$ is similar.

\noindent
{\bf Lemma 2} Consider the system (\ref{zhu1})-(\ref{rr7}) with parameters 
satisfying $\kappa<0$. Then there are choices of the 
remaining parameters for which there exist precisely two steady states 
and choices for which there do not exist any steady states. If,
on the other hand, $\kappa>0$ there are choices of the 
remaining parameters for which there exists precisely one steady state.

\noindent
{\bf Proof} Consider the case $\kappa<0$. Suppose first 
that $K_{m7}-K_{m4}>0$ and consider $p(\beta)$ for 
some fixed value of $\beta\in (0,1/5)$. Fix a value of $\frac{k_7}{k_4}$ and
a choice of all parameters except $k_4$ and $k_7$. 
Under the given assumptions the expression in the second line of (\ref{cubic}) 
is fixed. The expression in the first line is a fixed negative constant times
$k_4(\beta K_{m7}-(k_7/k_4)K_{m4})$. It follows directly from the definitions of
$v_4$ and $v_7$ that under the assumptions of the lemma the latter expression
is always positive. Thus the expression in the first line is negative and 
can be made arbitrarily large in magnitude by choosing $k_4$ large. Hence for 
large enough values of $k_4$ it can be concluded that $p(\beta)<0$ and it 
follows from the intermediate value theorem that $p$ has two roots in the 
interval $(0,1/5)$. Starting from one of these values of $\beta$ it is 
possible to define $x_{PGA}$ and $x_{GAP}$ by means of (\ref{pgaq}) and 
(\ref{gapq}). It has already been shown that the denominator of the right hand 
side of (\ref{gapq}) does not vanish and it follows in a similar way from the
definitions of $v_2$ and $v_6$ that the denominator of the right hand 
side of (\ref{pgaq}) does not vanish. This provides corresponding 
values of $v_2$, $v_4$, $v_6$ and $v_7$. By construction they satisfy 
(\ref{combmmz}) and (\ref{rational}). Hence they can be completed to a 
solution of (\ref{linearmmz}). If $k_1$, $k_3$ and $k_5$ are chosen 
sufficiently large the reaction rates can be reproduced by suitable values of 
$x_{\rm RuBP}$, $x_{\rm DPGA}$ and $x_{\rm Ru5P}$. Thus a steady state is 
obtained and this gives the first conclusion of the lemma. The inequality 
$\beta>\frac{k_7K_{m4}}{k_4K_{m7}}$ ensures that $\beta$ is bounded away from 
zero for all steady states. This implies a fixed positive lower bound 
for the expression in the second line of (\ref{cubic}). Hence this expression 
dominates the expression in the first line for $k_4$ small and the sum is 
positive, contradicting the existence of a root of $p$. This gives the second
conclusion of the lemma. The proofs in the case where $K_{m6}-K_{m21}>0$ are 
similar, with the role played by $k_4$ and $k_7$ in the preceding argument 
being taken over by $k_2$ and $k_6$.

Now consider the case $\kappa>0$. It was shown in the 
proof of Lemma 1 that in this case $p$ has precisely one root in the interval
$(0,1/5)$. The rest of the proof that there exists a steady state for
suitable choices of the other parameters is then as in the case where there
are two roots.

\noindent
{\bf Lemma 3}  Consider the system (\ref{zhu1})-(\ref{rr7}) with parameters 
satisfying $K_{m7}-K_{m4}=K_{m6}-K_{m21}=0$. If 
\begin{equation}\label{consistency}
\frac{k_6(x_{\rm ATP}+K_{m22})}{k_2x_{\rm ATP}}=\frac{1-5k_7/k_4}{5(1+k_7/k_4)}
\end{equation} 
then the steady states form a one-dimensional continuum. If 
(\ref{consistency}) does not hold then there are no positive steady states.

\noindent
{\bf Proof} When $K_{m7}-K_{m4}$ and $K_{m6}-K_{m21}$ are zero it follows as in
the proof of Lemma 1 that $\frac{v_7}{v_4}=\frac{k_7}{k_4}$ and 
$\frac{v_6}{v_2}=\frac{k_6(x_{\rm ATP}+K_{m22})}{k_2x_{\rm ATP}}$. The second 
part of the lemma then follows from (\ref{rational}). When (\ref{consistency})
holds we can proceed as follows. Choose $v_4$ to be an arbitrary number less
than $k_4$. Then define $v_7=\frac{k_7}{k_4}v_4$. Next let 
$v_6=\left(\frac15-\beta\right)v_4$ and define $v_2$ by the relation 
$v_6=\frac{k_6(x_{\rm ATP}+K_{m22})}{k_2x_{\rm ATP}}v_2$. These quantities can be
completed to a solution of (\ref{linearmmz}) and this in turn gives rise to a 
steady state of the evolution equations provided $k_1$, $k_3$ and $k_5$
are sufficiently large.

The main results which have been proved are summed up in 
the following theorem, which is a consequence of Lemma 1 - Lemma 3. 

\noindent
{\bf Theorem 1} Consider the system of \cite{zhu09} with all parameters 
positive.

\noindent
(i) There exist at most two isolated positive steady states.

\noindent
(ii) If there exist two isolated positive steady states then
$(K_{m7}-K_{m4})(K_{m6}-K_{m21})<0$ and if this inequality is satisfied then
there is an open region in the space of the other parameters for 
which two positive steady states exist.

\noindent
(iii) If $(K_{m7}-K_{m4})(K_{m6}-K_{m21})>0$ there is an open region in the space 
of the other parameters for which precisely one steady state exists. 

\noindent
(iv) There are values of the parameters where the quantities $K_{m7}-K_{m4}$
and $K_{m6}-K_{m21}$ are zero for which there exists a one-dimensional family of 
steady states. If either of these two quantities is non-zero then a 
continuum of steady states is not possible.

\noindent
{\bf Proof} (i) follows from Lemma 1 and Lemma 3. (ii) follows from Lemma 1 and
Lemma 2. (iii) follows from Lemma 2. (iv) follows from Lemma 1 and Lemma 3. 

It is instructive to compare the results of this theorem with the assertions
made in \cite{zhu09}. There fixed values are assumed for the Michaelis 
constants and this implies that $K_{m7}-K_{m4}=4.5$ and $K_{m6}-K_{m21}=0.51$,
putting us in the region where Theorem 1 implies that there exists at most
one positive steady state.

Note that in general if 
$L_1=x_{\rm RuBP}+\frac12 x_{\rm PGA}+\frac35 x_{\rm DPGA}
+\frac35 x_{\rm GAP}+x_{\rm Ru5P}$ then 
\begin{equation}
\frac{dL_1}{dt}=-\frac12 \left(v_6-\frac15 v_2\right)-\frac35 v_7.
\end{equation}
This is a Lyapunov function for $5v_6\ge v_2$. A sufficient condition for this
inequality to hold is that $K_{m6}\le K_{m21}$ and 
$5k_6\ge\frac{k_2x_{\rm ATP}}{x_{\rm ATP}+K_{m22}}$. It can be concluded that when
the parameters satisfy these inequalities there are no positive
steady states and other types of behaviour such as periodic solutions
can also be ruled out.

Next the stability of the steady states will be examined. The 
determinant of the derivative of the right hand side of the system 
(\ref{zhu1})-(\ref{zhu5}) is the sum of the product of the diagonal elements 
and the product of the off-diagonal elements. Thus it is a positive factor 
times
\begin{equation}\label{stabcrit}
-5\frac{\partial (v_2+v_6)}{\partial x_{\rm PGA}}
\frac{\partial (v_4+v_7)}{\partial x_{\rm GAP}}
+6\frac{\partial v_2}{\partial x_{\rm PGA}}\frac{\partial v_4}{\partial x_{\rm GAP}}.
\end{equation}
This expression can be manipulated further using the relation
$\frac{\partial v_4}{\partial x_{\rm PGA}}=\frac{K_{m4}v_4^2}{k_4x_{\rm PGA}^2}$
and the analogous expressions for the derivatives of $v_2$, $v_6$ and $v_7$. 
This leads to the following form of (\ref{stabcrit})
\begin{eqnarray}
&&\frac{1}{x_{\rm PGA}^2x_{\rm GAP}^2}
\left[-5\left(\frac{K_{m21}v_2^2(x_{\rm ATP}+K_{m22})}{k_2x_{\rm ATP}}
+\frac{K_{m6}v_6^2}{k_6}\right)
\left(\frac{K_{m4}v_4^2}{k_4}+\frac{K_{m7}v_7^2}{k_7}\right)\right.\nonumber\\
&&\left.+6\frac{K_{m21}v_2^2(x_{\rm ATP}+K_{m22})K_{m4}v_4^2}{k_2k_4x_{\rm ATP}}\right].
\end{eqnarray}
Dividing by the positive factor $v_2^2v_4^2$ and multiplying by 
$k_4x_{\rm PGA}^2x_{\rm GAP}^2$ leads to the expression
\begin{eqnarray}\label{betacrit}
&&-5\left(\frac{K_{m21}(x_{\rm ATP}+K_{m22})}{k_2x_{\rm ATP}}
+\frac{K_{m6}(1-5\beta)^2}{25k_6(1+\beta)^2}\right)
\left(K_{m4}+\frac{k_4K_{m7}\beta^2}{k_7}\right)\nonumber\\
&&+6\frac{K_{m21}K_{m4}(x_{ATP}+K_{m22})}{k_2x_{\rm ATP}}.
\end{eqnarray}
Without examining it in any detail we see that this expression has at most 
four zeroes for given values of the parameters and that in particular they
are isolated. 

All the coefficients in the characteristic polynomial other than the 
determinant are positive
and it is an increasing function for non-negative values of its argument
whose derivative at zero is positive. Thus either the linearization has 
precisely one positive eigenvalue or there is none and when there is a zero 
eigenvalue it is of multiplicity one. In general it is hard to determine the 
sign which distinguishes these two cases. Consider the case discussed in 
Lemma 2 where decreasing the value of $k_4$ takes us from a situation with two 
positive steady states to one with no positive steady states. There is a first 
value $\gamma$ of $k_4$ for which there are no longer two steady states and at 
that point there must be exactly one. At the parameter value $\gamma$ the 
linearization of the right hand side of the equations must have a non-trivial 
kernel. Thus in that case the characteristic polynomial has a root at zero
and this root is simple. Under suitable restrictions on the parameters all 
the other eigenvalues have negative real parts so that, in particular, there
are no eigenvalues which are purely imaginary and different from zero. This 
will be proved later. When this holds the centre manifold at that 
point is one-dimensional. (Some relevant background on centre manifold theory 
can be found in the Appendix.) The value of $\beta$ corresponding to this 
steady state must be a zero of (\ref{betacrit}). If $k_4$ is increased 
slightly there are two steady states. If (\ref{stabcrit}) held for one of 
these then it would have to have the same value of $\beta$ as the solution for 
$k_4=\gamma$. This can be true for at most one of the two solutions under 
consideration and so at least one of these has no zero eigenvalue. Consider 
now the perturbed centre manifold in the sense of the Appendix corresponding 
to a value of $k_4$ close to the critical value where there are two steady 
states. Let us fix an orientation of this one-dimensional manifold so that we 
can call one direction along it left and the other right. The sign of the 
vector field to the left of both steady states is the same as its sign to the 
right of both steady states. The vector field along this manifold changes sign 
at at least one of its two zeroes (where the relevant eigenvalue is non-zero) 
and hence also at the other. It follows that on the perturbed centre manifold 
one of the steady states is a sink and the other a source. If it could be 
proved that all the eigenvalues corresponding to eigenvectors transverse to 
the centre manifold have negative real parts then we would have shown that one 
of the steady states is asymptotically stable. In the other situation 
discussed in Lemma 2, where there is precisely one positive steady state, 
another approach can be used to obtain information about stability. In that 
case we consider the limit $k_6\to 0$. In the limit (\ref{stabcrit}) 
simplifies and shows that the sign of the critical quantity determining the 
sign of an eigenvalue of the linearization at the positive steady state is the 
same as that of $5k_7K_{m4}-k_4K_{m7}$. It follows by continuity that for $k_6$ 
small both signs occur for different values of the parameters.

Now the remaining four eigenvalues will be considered.
Start with a steady state and corresponding reaction rates $v_i$. Now
modify $k_3$ in such a way that it approaches $v_3$ from below. For each such 
value of $k_3$ the given value of $v_3$ can be produced by a unique value of 
$x_{\rm DPGA}$ and in this way
we get a one-parameter family of steady states. As $k_3$ approaches $v_3$
the concentration $x_{\rm DPGA}$ tends to infinity and the derivative
$\frac{\partial v_3}{\partial x_{\rm DPGA}}$ tends to zero. The other elements
of the linearization remain unchanged. Thus in the limit the matrix tends to
one with four negative eigenvalues and one eigenvalue zero. It follows that 
for values of $k_3$ close enough to the limit there are four eigenvalues with 
negative real parts. Information has now been obtained on the stability 
properties of some positive steady states. Next we consider the stability 
properties of the solution at the origin, which exists for all values of 
the parameters. In this case (\ref{stabcrit}) becomes
\begin{equation}
-5\left(\frac{K_{m2}(x_{\rm ATP}+K_{m22})}{k_2x_{\rm ATP}}
+\frac{K_{m6}}{k_6}\right)\left(\frac{K_{m4}}{k_4}+\frac{K_{m7}}{k_7}\right)
+6\frac{K_{m2}(x_{\rm ATP}+K_{m22})}{k_2x_{\rm ATP}}\frac{K_{m4}}{k_4}.
\end{equation}
It is again helpful to look at the limit $k_6\to 0$. Then the critical 
quantity determining the sign of an eigenvalue becomes
$k_4K_{m7}-5k_7K_{m4}$. The sign of the real parts of the four remaining 
eigenvalues can be controlled in the limit $k_3\to 0$. These observations
are summed up in the following theorem. 

\noindent
{\bf Theorem 2} Consider the system of \cite{zhu09} with all parameters
positive. 

\noindent
(i) There is an open set of parameter values for which there
exist one asymptotically stable and one unstable positive steady state.

\noindent
(ii) There is an open set of parameter values for which the unique positive
steady state is asymptotically stable and an open set for which the unique 
positive steady state is unstable.

\noindent
(iii) There is an open set of parameter values for which the origin is
asymptotically stable and an open set for which it is unstable.
 
\section{The Michaelis-Menten model of Grimbs et. al.}

In \cite{grimbs11} the authors introduced a variant of the model of \cite{zhu09}
where the stoichiometric coefficients are rescaled so as to make them all 
integers. The equations are \begin{eqnarray}
&&\frac{dx_{\rm RuBP}}{dt}=v_5-v_1,\label{grimbs1}\\
&&\frac{dx_{\rm PGA}}{dt}=2v_1-v_2-v_6,\label{grimbs2}\\
&&\frac{dx_{\rm DPGA}}{dt}=v_2-v_3,\label{grimbs3}\\
&&\frac{dx_{\rm GAP}}{dt}=v_3-5v_4-v_7,\label{grimbs4}\\
&&\frac{dx_{\rm Ru5P}}{dt}=3v_4-v_5.\label{grimbs5}
\end{eqnarray} 
Different models were considered in \cite{grimbs11} with different kinetics.
In one of these, which is studied in this section, Michaelis-Menten kinetics
is used. Although it is not stated explicitly in \cite{grimbs11} we assume
that the model is identical to that of \cite{zhu09} except for the modified
stoichiometric coefficients. In other words, the $v_i$ are defined as in
equations (\ref{rr1})-(\ref{rr7}) in the previous section except that 
$x_{\rm GAP}$ is replaced by $x_{\rm GAP}^5$ in the expression for $v_4$. For short 
we call this the MM model. In another
model discussed in \cite{grimbs11} each of the basic reactions is replaced
by a Michaelis-Menten scheme with substrate, enzyme and substrate-enzyme 
complex and the elementary reactions are given mass action kinetics. 
Following \cite{grimbs11} we call this the MM-MA model (Michaelis-Menten via 
mass action). As mentioned in \cite{rendall14} steady states of the MM
model are in one-to-one correspondence with steady states of the 
MM-MA model with fixed total amounts of substrates and enzymes. It was shown
in \cite{rendall14} that for suitable values of the parameters the MM-MA model
has more than one positive steady state with the same total amounts
of substrates and enzymes. It follows that for suitable values of the 
parameters the MM model has  more than one positive steady state.
In \cite{rendall14} no information was obtained on the stability of the 
steady states. In what follows it will be shown that there
are parameter values for which one of the positive steady states is 
stable and the other is unstable.

We now investigate steady states of the MM system directly. There
is a calculation for the reaction rates analogous to that in the last 
section. For a steady state $v_1=v_5$, $2v_1=v_2+v_6$, $v_2=v_3$, 
$v_3=5v_4+v_7$ and $v_5=3v_4$. Combining these gives 
\begin{equation}\label{combmm}
0=v_3-5v_4-v_7=v_2-5v_4-v_7=2v_1-5v_4-v_6-v_7=v_4-v_6-v_7.
\end{equation} 
It follows that if $\beta=\frac{v_7}{v_4}$ then 
\begin{equation}\label{rationalmm}
\frac{v_6}{v_2}=\frac{1-\beta}{5+\beta}
\end{equation}
For a positive steady state we must have $0<\beta<1$. It can be checked that
the equations for steady states are equivalent to (\ref{combmm}), 
(\ref{rationalmm}) and the equations
\begin{equation}\label{remainingmm}
v_5=3v_4, v_3=5v_4+v_7, v_1=3v_4.
\end{equation} 
Thus we see that as in the last section solutions of (\ref{combmm}) and 
(\ref{rationalmm}) can be completed to steady states of the whole system.
An analogous statement holds in the limiting case $k_6=0$, where $v_6=0$,
$v_4=v_7$ and $\beta=1$. Then the equations (\ref{remainingmm}) must
be complemented by the equation $v_2=v_3$.

In the case of the MM system the route to analysing steady states used in the 
previous section does not appear useful. It is possible to express 
$\frac{v_7}{v_4}$ as a function of $x_{\rm GAP}$ but this relation cannot be 
solved for $x_{\rm GAP}$. For this reason we now concentrate on the limiting 
case $k_6=0$ where we have $v_4=v_7$. Some information on the case $k_6\ne 0$ 
will be obtained later. In the case $k_6=0$ the main equation to be solved is
\begin{equation}\label{balance}
\frac{k_4x_{\rm GAP}^4}{K_{m4}+x_{\rm GAP}^5}=\frac{k_7}{K_{m7}+x_{\rm GAP}}.
\end{equation}
Rearranging gives
\begin{equation}
(k_4-k_7)x_{GAP}^5+k_4K_{m7}x_{GAP}^4=k_7K_{m4}.
\end{equation}
It turns out (cf. \cite{rendall14}, Lemma 2) that this equation has two 
positive solutions precisely when $k_4<k_7$ and 
$\frac15 k_4K_{m7}\left[\frac{4k_4K_{m7}}{5(k_7-k_4)}\right]^4<k_7K_{m4}$. 
There is a bifurcation when 
$k_7-k_4=\frac15\left[\frac{(\frac15 k_4K_{m7})^5}{k_7K_{m4}}\right]^{\frac14}$.
This happens exactly when the derivatives of the terms on both sides of 
(\ref{balance}) are equal. Once (\ref{balance}) has been solved for $x_{\rm GAP}$
this can be completed to a steady state as explained above.

When $k_6\ne 0$ things are more complicated. One tractable special case is that
where $K_{m21}=K_{m6}$. Then we get 
\begin{equation}
v_4\left[1-6\left(\frac{k_6}{k_2+k_6}\right)\right]=v_7.
\end{equation}
Provided $k_2>5k_6$ the analysis of this relation is just as in the case 
$k_6=0$ except for the fact that $k_4$ is replaced by 
$\left[1-6\left(\frac{k_6}{k_2+k_6}\right)\right]k_4$.
Once $x_{\rm GAP}$ has been determined for a steady state it is possible 
to reconstruct the concentrations of Ru5P, RuBP, PGA and DPGA.

Starting from the result in the case $k_6=0$ we can obtain a result for $k_6$
small but non-zero using the implicit function theorem. The reaction rates
satisfy the equations $v_4=v_6+v_7$ and 
$\frac{v_6}{v_2}=\frac{1-\frac{v_7}{v_4}}{5+\frac{v_7}{v_4}}$. Substituting the
definitions of the reaction rates into these gives two equations for the 
concentrations $x_{\rm PGA}$ and $x_{\rm GAP}$. When $k_6=0$ the first of these
equations simplifies to the equation for $x_{\rm GAP}$ alone which has just 
been analysed. When it has two solutions the derivative of $v_4-v_6-v_7$ with 
respect to $x_{\rm GAP}$ is non-zero at each of these. To show that these
solutions persist for $k_6$ small using the implicit function theorem it 
suffices to show that the partial derivative of $\frac{v_6}{v_2}$ with respect
to $x_{\rm PGA}$ is non-zero for $k_6=0$. This is easily checked. Since it
was shown in \cite{rendall14} that there are parameter values for which
there exists a continuum of solutions of the MM-MA system with the same 
values of the conserved quantities there must also exist parameters for
which the MM system has a continuum of solutions but solutions of this type
will not be studied further here.

Next the stability of the steady states will be investigated by looking
at the derivative of the right hand side of the equations. Consider first
the case $k_6=0$. The constant term in the characteristic polynomial is the 
product of a positive quantity with 
\begin{equation}
\frac{5k_4K_{m4}x_{\rm GAP}^4}{(K_{m4}+x_{\rm GAP}^5)^2}
-\frac{k_7K_{m7}}{(K_{m7}+x_{\rm GAP})^2}
\end{equation}
This vanishes precisely when the bifurcation condition holds. The characteristic
polynomial has properties analogous to those we saw in the last section. When 
the characteristic polynomial vanishes at zero it has a simple root there. 
Its derivative at that point is non-zero. There are no other positive real 
roots. It will be shown later that there are parameter values for which there
exist no eigenvalues which are purely imaginary but different from zero. When 
this holds the centre manifold at the bifurcation point is one-dimensional
and we have a situation similar to that in the last section. When there are two 
steady states on the perturbed centre manifold both of them are 
hyperbolic. When $k_6$ is perturbed a little away from zero these two solutions
continue to exist and to be hyperbolic. That it can be arranged that four of 
the eigenvalues have non-zero (in fact negative) real parts can be shown by the 
same method as in the previous section, considering the limit $k_3\to 0$.
It can be seen that one of the steady states is a hyperbolic sink. The 
unstable manifold of the other steady state coincides with part of the 
perturbed centre manifold and is a heteroclinic orbit connecting the two
steady states. 

In \cite{rendall14} it was shown how a parameter $\epsilon$ can be introduced
into the MM-MA system so as to obtain the MM system formally in the limit
$\epsilon\to 0$. This was done for a more general class of systems including
the MM-MA system for the Calvin cycle as a special case. In fact this is more
than a formal limit and rigorous results on stability can be obtained using
geometric singular perturbation theory (GSPT). (A basic reference for this
subject is \cite{fenichel79} and a summary of some of the key ideas can be
found in the Appendix of \cite{hell15b}.) To achieve this it is necessary
to show that certain eigenvalues, the transverse eigenvalues in the sense of 
GSPT, have non-zero real parts. In the example considered here we will show that
they all have negative real parts. Under these circumstances we can say the 
following. If there is a hyperbolic steady state of the MM system, $k$
of whose eigenvalues have negative real parts then for $\epsilon$ small the 
corresponding steady state of the MM-MA system is hyperbolic and 
has $k+t$ eigenvalues with negative real parts, where $t$ is the number of
transverse eigenvalues. In particular, if the solution of the MM system is
stable the same is true for the solution of the MM-MA system.  

In the general notation used in \cite{rendall14} the substrates are denoted
by ${\rm A_i}$ and in the present example they are the five substances 
occurring in the MM model. The enzymes catalysing the seven reactions are 
denoted by ${\rm E_\alpha}$. The complex formed by the binding of ${\rm A_i}$ 
to ${\rm E_\alpha}$ is denoted by ${\rm A_iE_\alpha}$. After a suitable rescaling 
the MM-MA system takes the form 
\begin{eqnarray}
&&\dot x=f(x,y,\epsilon),\\
&&\epsilon\dot y=g(x,y,\epsilon),
\end{eqnarray}  
which is the standard form used in GSPT. In the example $x$ consists of the
variables $x_{\rm A_i}$, $y$ consists of the variables $x_{\rm A_iE_\alpha}$ and the 
variables $x_{\rm E_\alpha}$ have been eliminated using the conservation laws for 
the total amounts of enzymes. The
transverse eigenvalues are the eigenvalues of the derivative of the right hand 
side of the equation for $y$ with respect to the variable $y$. In this type of
system the evolution equations for the different substrate-enzyme complexes are
all decoupled from each other since each enzyme only binds to one substrate.
Hence the matrix whose eigenvalues are to be calculated is diagonal. The 
eigenvalues can be read off from the equations in \cite{rendall14}. For 
example in the notation of that paper the eigenvalue corresponding to the 
variable $x_{\rm GAPE_4}$ is $-k_{10}x_{\rm GAP E_4}^5-k_{11}-k_{12}$. It can be 
concluded
that in the case $k_6=0$ there exist parameter values for which the MM-MA system
has two hyperbolic steady states, one of which is asymptotically 
stable and the other of which has a one-dimensional unstable manifold. By 
continuity the same holds for $k_6$ small and non-zero.  Moreover the stable
and unstable steady states are connected by a heteroclinic orbit.

\noindent
{\bf Theorem 3} There are positive parameter values for the MM system for
which there exist one stable and one unstable positive steady state. The
same holds for the MM-MA system with suitable fixed values of the total 
amounts of substrates and enzymes.

\section{The MAdh model}

A model for the Calvin cycle including diffusion was introduced in
\cite{grimbs11}. It uses mass action kinetics.
Restricting consideration to spatially homogeneous solutions
or setting the diffusion constant to zero leads to a system of six ordinary
differential equations which was studied in \cite{rendall14} and was called
the MAdh model. It was shown that for certain values of the parameters there 
exist two positive steady states. The stability of those solutions was 
not determined. Here we will show how information about their stability can be 
obtained. In the MAdh model equations (\ref{grimbs1})-(\ref{grimbs5}) hold
and are supplemented by the equation
\begin{equation}
\frac{dx_{\rm ATP}}{dt}=-v_2-v_5+v_8\label{madh}.
\end{equation}
The reaction rates are
\begin{eqnarray}
&&v_1=k_1x_{\rm RuBP},\label{rrma1}\\
&&v_2=k_2x_{\rm PGA}x_{\rm ATP},\label{rrma2}\\
&&v_3=k_3x_{\rm DPGA},\label{rrma3}\\
&&v_4=k_4x_{\rm GAP}^5,\label{rrma4}\\
&&v_5=k_5x_{\rm Ru5P}x_{\rm ATP},\label{rrma5}\\
&&v_6=k_6x_{\rm PGA},\label{rrma6}\\
&&v_7=k_7x_{\rm GAP},\label{rrma7}\\
&&v_8=k_8(c-x_{\rm ATP})\label{rrma8}
\end{eqnarray}
with a positive constant $c$. The equations for the $v_i$ with $1\le i\le 7$ 
satisfied by steady states of the MM system are also valid for the MAdh 
system. There is an extra reaction rate $v_8$ for the regeneration of ATP and 
an extra equation $v_8=v_2+v_5$ which can be solved at the end if required. 

In \cite{rendall14} is was shown that positive steady states of the 
MAdh model are only possible if $x_{\rm ATP}>\frac{5k_6}{k_2}$ and that
determining them is equivalent to finding solutions of the following system of 
two equations for the concentrations $x_{\rm GAP}$ and $x_{\rm ATP}$. 
\begin{eqnarray}
&&f_1(x_{\rm GAP},x_{\rm ATP})=k_4(k_2x_{\rm ATP}-5k_6)x_{\rm GAP}^4
-k_7(k_2x_{\rm ATP}+k_6)=0,\label{madhs1}\\
&&f_2(x_{\rm GAP},x_{\rm ATP})=x_{\rm ATP}-c+\frac{8k_4}{k_8}x_{\rm GAP}^5
+\frac{k_7}{k_8}x_{\rm GAP}=0.\label{madhs2}
\end{eqnarray}
A solution of these equations can be completed to a steady state of
the whole system by defining
\begin{eqnarray}
&&x_{\rm RuBP}=\frac{3k_4}{k_1}x_{\rm GAP}^5,\\
&&x_{\rm PGA}=\frac{2k_1x_{\rm RuBP}}{k_2x_{\rm ATP}+k_6},\\
&&x_{\rm DPGA}=\frac{k_7x_{\rm GAP}}{k_3}+\frac{5k_4x_{\rm GAP}^5}{k_3},\\
&&x_{\rm Ru5P}=\frac{3k_4x_{\rm GAP}^5}{k_5x_{\rm ATP}}.
\end{eqnarray}
Moreover, depending on the parameters the number of solutions of the equations 
(\ref{madhs1}) and (\ref{madhs2}) is zero, one or two. We call the values of 
the parameters for which there is exactly one solution the bifurcation values.
They are precisely the points where the Jacobian determinant of the mapping
$(f_1,f_2)$ vanishes. It is clear that the zero sets of $f_1$ and $f_2$ are
smooth curves. A bifurcation point occurs precisely when these two curves are 
tangent to each other.

We claim that the characteristic polynomial has a zero eigenvalue at the 
bifurcation point. This is because as $c$ is varied while the other parameters
are fixed two steady states coalesce at the bifurcation point. We will
show later that there are parameter values for which this zero has 
multiplicity one. To do this we study the linearization of the right hand side
of the equations. To compute its eigenvalues it is necessary to calculate
the determinant of a certain matrix. Adding suitable multiples of the second 
and fifth columns of this matrix to the last column simplifies the matrix while
leaving its determinant unchanged. The determinant of the linearization at a
steady state is a positive multiple of
\begin{eqnarray}\label{f1f3}
&&k_8\left[\left(1+\frac{k_6}{k_2x_{\rm ATP}}\right)
(25k_4x_{\rm GAP}^4+k_7)-30k_4x_{\rm GAP}^4\right]\nonumber\\
&&+\frac{6k_4k_6x_{\rm GAP}^5}{x_{\rm ATP}(k_2x_{\rm ATP}+k_6)}
(40k_4x_{\rm GAP}^4+k_7).
\end{eqnarray}
Denote this function of $x_{\rm GAP}$ and $x_{\rm ATP}$ by $f_3$. Note that it 
does not depend on $c$. Its partial derivative with respect to $x_{\rm ATP}$ is 
everywhere negative. Thus the zero set of $f_3$ is a smooth curve. For 
parameter values corresponding to a bifurcation point the zero sets of $f_1$ 
and $f_2$ are tangent to each other and the zero set of $f_3$ must pass 
through their point of tangency. If $c$ is increased the intersection
point splits into two and these two points lie in the
zero set of $f_1$. We would like to show that neither of these two points can 
lie in the zero set of $f_3$ if the parameters are sufficiently close to their 
bifurcation value. For in that case if the zero eigenvalue at the 
bifurcation point is of multiplicity one the linearization has no zero 
eigenvalues after bifurcation and the two steady states are hyperbolic. To 
get this conclusion it suffices to show that in a neighbourhood of the 
bifurcation point the zero sets of $f_1$ and $f_3$ intersect in only one
point. The equation $f_1=0$ can be used to solve for $x_{\rm ATP}$. Substituting
this into the expression for the determinant and multiplying by a suitable 
positive quantity gives a polynomial equation for the value of $x_{\rm GAP}$
at an intersection of the zero sets of $f_1$ and $f_3$. We know that this
polynomial vanishes at the bifurcation point and provided it does not
vanish identically its zero set is discrete, which gives the desired
result. That the polynomial does not vanish identically follows from the 
fact under the condition $f_1=0$ the expression (\ref{f1f3}) diverges as 
$x_{\rm GAP}$ tends to infinity. For in that situation $x_{\rm ATP}$ tends to 
a constant value and the second term in (\ref{f1f3}) dominates the first.

To show that there are parameter values for which the zero eigenvalue at the 
bifurcation point does indeed have multiplicity one we look at the limit 
where $k_5$ tends to zero while the other parameters are held constant.
The linearization then tends to a limit which is simpler than it is in general.
In the limit it can be easily seen than there are eigenvalues zero, 
$-k_7-25k_4x_{\rm GAP}^4$, $-k_3$ and $-k_1$. It remains to study the 
determinant
of a $2\times 2$ matrix. These are the roots of the quadratic polynomial
\begin{equation}
\lambda^2+(k_2x_{\rm ATP}+k_6+k_8+k_2x_{\rm PGA})\lambda+(k_2x_{\rm ATP}+k_6)k_8
+k_2k_6x_{\rm PGA}.
\end{equation}
They have negative real parts. Thus when $k_5$ is close to but not equal to
zero the bifurcation point is such that all eigenvalues except one have 
negative real parts. The sign of the remaining eigenvalue is then the same
as that of the determinant. In this situation the stable steady state is 
that at which the concentration of ATP is higher. Putting these facts together 
leads to the following result.
 
\noindent
{\bf Theorem 4} There exist positive parameter values for the MAdh system for
which there exist one stable and one unstable positive steady state.  

\section{Summary and outlook}

There are a number of things which have been proved about the dynamics
of simple models of the Calvin cycle where the unknowns are the 
concentrations of five sugar phosphates. In the present paper and the 
previous work on which it builds information has been obtained on the number 
and stability of positive steady states under various assumptions as well as
solutions where the concentrations tend to zero or infinity at late times.
Similar information has been obtained on the MM-MA model and the MAdh 
model described in the paper. It is known that $\omega$-limit points
for which some concentrations are zero must be such that the concentrations 
of all sugar phosphates vanish \cite{rendall14}. All these models 
except the MM-MA and MAdh models are cooperative systems. This means that all 
the off-diagonal elements of the derivative of the right hand side of the 
equations are non-negative. In addition the derivative is irreducible, i.e. it 
leaves no
linear subspace defined by the vanishing of a subset of the variables invariant.
Thus, by a theorem of Hirsch \cite{hirsch85}, in the set of initial data giving
rise to bounded solutions all but those belonging to a set of measure zero
converge to the set of steady states at late times. When the steady states are 
isolated this means that each of these solutions converges to a steady state. 
On the other hand it is not known whether every bounded solution converges to 
a steady state and it is also not known whether there exist periodic solutions.
Furthermore, the maximum number of isolated steady states for a given model 
is generally not known. For the model of \cite{zhu09} and the MAdh model
there are never more than two of these. For the MM model this question is 
still open and it is not clear how it could be approached.

There are many other models of the Calvin cycle and in general they include 
many chemical species. An interesting model was introduced in 
\cite{pettersson88} and a modified version of it was studied in 
\cite{poolman99} and \cite{poolman00}.
In these models the kinetics is more complicated than mass action or 
Michaelis-Menten with the concentrations of some species modulating the 
rate of reactions where they are not among the reactants. These models and 
simplified variants of them with mass action kinetics have been investigated
mathematically in \cite{moehring15}. Information was obtained about solutions
for which some concentrations tend to zero at late times. This is related to
the biological phenomenon known as overload breakdown. It is also related to
the cases in the present paper where concentrations tend to zero at late times.
There remains much to be understood concerning the dynamics of these models.
It would be desirable to understand the mathematical relation of these models
to the models including less chemical species.

The techniques used in this paper might also be applied to other problems. 
Here it was seen that when the centre manifold at the bifurcation is of
dimension one it may be possible to obtain results about stability on
the basis of quite limited information. This may be compared with the example in
\cite{hell15b} of a model of a biochemical system where the method used to 
control a bifurcation with a one-dimensional centre manifold made use of a lot 
more detailed calculations. Another technique which played a central role was 
looking at limiting values of the parameters to get information about what 
types of stability properties can occur. In many arguments it turned out to be 
useful to  carry the calculations as far as possible in terms of the reaction 
rates before using the dependence of these rates on the concentrations.

\section*{Appendix: some background on centre manifolds}

Here some facts about centre manifolds which are relevant to the paper are
reviewed briefly. Consider a system of ODE of the form $\dot x=f(x)$ and a 
steady state $x_0$. In other words $f(x_0)=0$. In general the derivative 
$Df(x_0)$ has eigenvalues with positive, zero and negative real parts. The 
corresponding generalized eigenvectors define three linear subspaces 
$E_u$, $E_c$ and $E_s$ whose direct sum is the whole space. $E_c$ is the centre 
subspace. There exists a manifold $V_c$ (in general non-unique) called the 
centre manifold of $x_0$, which contains $x_0$, is invariant under the evolution
defined by the ODE and whose tangent space at $x_0$ is $E_c$. The dynamics 
close to $x_0$ is topologically equivalent to a product of two factors. One
factor is the dynamics on any centre manifold of $x_0$. The other is 
topologically equivalent to a linear system with $E_u$ and $E_s$ of the 
same dimensions as in the original system. This means that if we understand 
the qualitative properties of solutions which are close to $x_0$ and lie on 
the centre manifold we obtain information on the dynamics of all solutions 
close to $x_0$. All steady states sufficiently close to $x_0$ lie on 
the centre manifold of $x_0$. For more information on these matters we refer 
to \cite{kuznetsov10}, in particular Chapter 5 of that book.

Next suppose that we have instead of a single ODE a family 
$\dot x=f(x,\lambda)$ of ODE depending on a parameter $\lambda$ and suppose 
that $(x_0,0)$ is a steady state. An extended system with one more dimension
can be defined by adjoining the equation $\dot\lambda=0$. The centre manifold
of the extended system at $(x_0,0)$ has one more dimension than that of the
centre manifold of the original system at $x_0$. Since $\lambda$ is time 
independent the extended centre manifold is foliated by invariant manifolds
of constant $\lambda$ which agree with the original centre manifold for 
$\lambda=0$. Let us call a manifold of this type a perturbed centre manifold.
This construction is useful for the study of the case where $(x_0,0)$ is a 
bifurcation point, i.e. the centre manifold for the system with $\lambda=0$ at 
$x_0$ has dimension greater than zero. In this paper we are concerned with the 
case that the dimension of the centre manifold is one so that the dynamics on 
these invariant manifolds is of dimension one. For $(x,\lambda)$ sufficiently
close to $(x_0,0)$ all steady states of the system for a fixed value 
of $\lambda$ lie on the perturbed centre manifold corresponding to that value
of the parameter.

\end{document}